\documentclass[12pt]{article}
\usepackage{amsmath,amsfonts,latexsym,amsthm,amssymb,mathabx}
\newcommand{\D}{\mathbb D_t^{(\alpha )}}
\newcommand{\Rn}{\mathbb R^n}

\DeclareMathOperator{\R}{Re}
\DeclareMathOperator{\I}{Im}
\newcommand{\DK}{\mathbb D_{(k)}}
\newcommand{\A}{\widetilde{A}}
\newcommand{\B}{\widetilde{B}}
\newcommand{\vk}{\varkappa}
\newcommand{\K}{\mathcal K}
\newcommand{\LL}{\mathcal L}
\newcommand{\CBF}{\mathcal C\mathcal B\mathcal F}
\begin{document}
\newtheorem{prop}{Proposition}[section]
\newtheorem{lem}{Lemma}[section]
\newtheorem{teo}{Theorem}[section]
\pagestyle{plain}
\title{ Fractional kinetic hierarchies and intermittency}

\author{ \textbf{Anatoly N.
Kochubei}\\
Institute of Mathematics,\\
National Academy of Sciences of Ukraine, \\
Tereshchenkivska 3, \\
Kyiv, 01601 Ukraine\\
Email: kochubei@imath.kiev.ua
\and
\textbf{Yuri Kondratiev}\\
Department of Mathematics, University of Bielefeld, \\
D-33615 Bielefeld, Germany,\\
Email: kondrat@math.uni-bielefeld.de}

\numberwithin{equation}{section}
\date{}
\maketitle

\begin{abstract}

We consider general convolutional  derivatives
and related fractional statistical dynamics of continuous interacting particle systems.
We apply the subordination principle to  construct
kinetic fractional statistical dynamics in the continuum in terms of  solutions   to
Vlasov-type
hierarchies.   Conditions for the intermittency  property of fractional kinetic  dynamics are obtained.
\medskip{}

\end{abstract}

\section{Introduction}

Kinetic equations for classical gases may be derived from the BBGKY hierarchies for time dependent correlation functions
which describe Hamiltonian dynamics of gases,
see e.g. an excellent review by H.Spohn \cite{Spohn80}.  Making  scalings in BBGKY hierarchical chains, we will arrive
in the limiting kinetic hierarchies of Boltzmann or Vlasov type  depending on the particular scaling we use. Both kinetic hierarchies have a common chaos propagation property. Using this property we obtain Boltzmann or Vlasov equation
respectively as non-linear equations for the density of the considered system.

A similar approach may be also applied to Markov dynamics of interacting particle systems in the continuum as it was proposed in
\cite{FKK10}.  These dynamics may be described on the microscopic level by means of related hierarchical evolution equations for
correlation functions and proper scalings will lead to  limiting  mesoscopic hierarchies and corresponding kinetic equations.
Again, a common point for  resulting  hierarchies is the chaos propagation
 property that is a root of the kinetic equation for the density of the system. Note that this property means that the kinetic state evolution of the system will be given by a flow of Poisson measures provided the initial state is a Poisson measure.
Of course, a rigorous realization of this scheme  (that includes such steps as  construction of the microscopic Markov dynamics,
 control of the convergence of  solutions  for rescaled evolutions and an analysis of corresponding kinetic equations) shall be
 done for each particular model and is, in general, quite difficult technical problem. At the present time, this program is realized for a number of Markov dynamics of continuous systems which includes certain birth-and-death processes, Kawasaki type dynamics, binary jumps models, see e.g.  \cite{FKK10, FKK12, FKKK15}.

 In the present paper we extend described above approach to the case of certain non-Markov dynamics of interacting
 particle systems in the continuum. Namely, we will consider hierarchical evolution equations for correlation functions with
general  fractional time derivatives of convolutional type.   From the stochastic point of view,  the latter corresponds to a random time change
 in the original Markov processes and effectively leads to a memory effect in stochastic dynamics. The Vlasov type mesoscopic scaling for
 the fractional hierarchical chains will affect only spatial structure of their  generators and will give  kinetic hierarchies of the same form as before but with fractional time derivatives. The latter drastically change the structure of  their solutions.  In terms of corresponding state evolutions we obtain subordinations of Poisson flows. The latter means that in the fractional case the kinetic hierarchies are not reduced just to density evolutions. Time development of correlation functions in such hierarchical chains is essentially different for all levels of the hierarchy. In other words, the kinetic description of the dynamics needs to work with all the hierarchy but not only with the evolution of the density.  As a very prominent effect of this situation we will show an intermittency
 property for certain classes of fractional kinetic dynamics. This property means a progressive growth in the time for correlation functions of higher orders and never may be observed for Poisson flows. Note that for the classical case of Caputo-Djrbashian
 fractional derivatives we pointed out this effect in our previous paper \cite{DaSilvaKochKon}. Here we are dealing with a large class
 of generalized time derivatives including, in particular,  the case of so-called distributed order derivatives.

\section{General fractional calculus}

\medskip
{\bf 2.1.} In order to explain a general concept of fractional calculus developed in \cite{K11}, we need some notions from function theory connected with properties of the Laplace transform. For their detailed exposition see \cite{SSV}.

A real-valued function $f$ on $(0,\infty )$ is called a {\it Bernstein function}, if $f\in C^\infty$, $f(\lambda )\ge 0$ for all $\lambda >0$, and
$$
(-1)^{n-1}f^{(n)}(\lambda )\ge 0\quad \text{for all $n\ge 1,\lambda >0$},
$$
so that its derivative $g=f'$ is completely monotone, that is $(-1)^mg^{(m)}(\lambda )\ge 0$, $m=0,1,2,\ldots$.

Equivalently, a function $f:\ (0,\infty )\to \mathbb R$ is a Bernstein function, if and only if
\begin{equation}
\label{3.1}
f(\lambda )=a+b\lambda +\int\limits_0^\infty \left( 1-e^{-\lambda t}\right)\,\mu (dt)
\end{equation}
where $a,b\ge 0$, and $\mu$ is a Borel measure on $[0,\infty )$, called the {\it L\'evy measure}, such that
\begin{equation}
\label{3.2}
\int\limits_0^\infty \min (1,t)\,\mu (dt)<\infty .
\end{equation}
The triplet $(a,b,\mu )$ is determined by $f$ uniquely. In particular,
\begin{equation}
\label{3.3}
a=f(0+),\quad b=\lim\limits_{\lambda \to \infty}\frac{f(\lambda )}{\lambda}.
\end{equation}

A Bernstein function $f$ is said to be a {\it complete Bernstein function}, if its L\'evy measure $\mu$ has a completely monotone density $m(t)$ with respect to the Lebesgue measure, so that (\ref{3.1}) takes the form
\begin{equation}
\label{3.4}
f(\lambda )=a+b\lambda +\int\limits_0^\infty \left( 1-e^{-\lambda t}\right) m(t)\,dt
\end{equation}
where, by (\ref{3.2}),
\begin{equation*}
\int\limits_0^\infty \min (1,t)m(t)\,dt<\infty .
\end{equation*}
Here the complete monotonicity means that $m\in C^\infty (0,\infty )$, $(-1)^nm^{(n)}(t)\ge 0$, $t>0$, for all $n=0,1,2,\ldots$.

Another important class of functions is that of {\it Stieltjes functions}, that is of functions $\varphi$ admitting the integral representation
\begin{equation}
\label{3.5}
\varphi (\lambda )=\frac{a}{\lambda}+b +\int\limits_0^\infty \frac1{\lambda +t}\sigma (dt)
\end{equation}
where $a,b\ge 0$, $\sigma$ is a Borel measure on $[0,\infty )$, such that
\begin{equation}
\label{3.6}
\int\limits_0^\infty (1+t)^{-1}\sigma (dt)<\infty .
\end{equation}

Using the identity $(\lambda +t)^{-1}=\int\limits_0^\infty e^{-ts}e^{-\lambda s}ds$ we find from (\ref{3.5}) that
\begin{equation}
\label{3.7}
\varphi (\lambda )=\frac{a}{\lambda}+b +\int\limits_0^\infty e^{-\lambda s}g(s)\,ds
\end{equation}
where
\begin{equation}
\label{3.8}
g(s)=\int\limits_0^\infty e^{-ts}\sigma (dt)
\end{equation}
is a completely monotone function whose Laplace transform exists for any $\lambda >0$.

We will denote the class of complete Bernstein functions by $\CBF$, and the class of Stieltjes functions by $\mathcal S$. The following characterization is proved in \cite{SSV}: for a nonnegative function $f$ on $(0,\infty )$, the following conditions are equivalent.

\begin{description}
\item[(i)] $f\in \CBF$.

\item[(ii)] The function $\lambda \mapsto \lambda^{-1}f(\lambda )$ is in $\mathcal S$.

\item[(iii)] $f$ has an analytic continuation to the upper half-plane $\mathbb H=\{z\in \mathbb C:\ \I z>0\}$, such that $\I f(z)\ge 0$ for all $z\in \mathbb H$, and there exists the real limit
    \begin{equation}
    \label{3.9}
    f(0+)=\lim\limits_{(0,\infty )\ni \lambda \to 0}f(\lambda ).
\end{equation}

\item[(iv)] $f$ has an analytic continuation to the cut complex plane $\mathbb C\setminus (-\infty ,0]$, such that $\I z\cdot \I f(z)\ge 0$, and there exists the real limit (\ref{3.9}).

\item[(v)] $f$ has an analytic continuation to $\mathbb H$ given by the expression
    \begin{equation}
    \label{3.10}
    f(z)=a+bz+\int\limits_0^\infty \frac{z}{z+t}\sigma (dt)
    \end{equation}
    where $a,b\ge 0$, and $\sigma$ is a Borel measure on $(0,\infty )$ satisfying (\ref{3.6}).
\end{description}

\bigskip
Note that the constants $a,b$ are the same in both the representations (\ref{3.4}) and (\ref{3.10}). The density $m(t)$ appearing in the integral representation (\ref{3.4}) of a function $f\in \CBF$ and the measure $\sigma$ corresponding to the Stieltjes function $\varphi (\lambda )=\lambda^{-1}f(\lambda )$ are connected by the relation
\begin{equation*}
m(t)=\int\limits_0^\infty e^{-ts}s\,\sigma (ds).
\end{equation*}

The importance of complete Bernstein functions is caused by the following ``nonlinear'' properties \cite{SSV} having significant applications.

\medskip
\begin{prop}
\begin{description}
\item[(i)] A function $f\not\equiv 0$ is a complete Bernstein function, if and only if $1/f$ is a Stieltjes function.

\item[(ii)] Let $f,f_1,f_2\in \CBF$, $\varphi ,\varphi_1,\varphi_2 \in \mathcal S$. Then $f\circ \varphi \in \mathcal S$, $\varphi \circ f\in \mathcal S$, $f_1\circ f_2\in \CBF$, $\varphi_1\circ \varphi_2 \in \CBF$, $(\lambda +f)^{-1}\in \mathcal S$ for any $\lambda >0$.
\end{description}
\end{prop}

\medskip
{\bf 2.2.} In fractional evolution equations, instead of the first time derivative, one considers nonlocal integro-differential operators. The simplest example of such an operator, for which a well-posed Cauchy problem is formulated as for the first order equations, is the Caputo-Djrbashian fractional derivative
\begin{equation}
\label{3.11}
\left( \mathbb D^{(\alpha )}u\right) (t)=\frac{1}{\Gamma
(1-\alpha )}\left[ \frac{d}{dt}\int\limits_0^t(t-\tau )^{-\alpha
}u(\tau )\,d\tau -t^{-\alpha }u(0)\right],\quad
t>0,
\end{equation}
where $0<\alpha <1$. For further details see, for example, \cite{KST}.

More generally, it is natural to consider differential-convolution operators
\begin{equation}
\label{3.12}
(\DK u)(t)=\frac{d}{dt}\int\limits_0^tk(t-\tau )u(\tau )\,d\tau -k(t)u(0)
\end{equation}
where $k\in L_1^{\text{loc}}(\mathbb R_+)$ is a nonnegative function.

A nontrivial example of an operator (\ref{3.12}) is a distributed order derivative $\mathbb D^{(\mu )}$ corresponding to
\begin{equation}
\label{3.13}
k(t)=\int\limits_0^1\frac{t^{-\alpha }}{\Gamma (1-\alpha )}\mu
(\alpha )\,d\alpha ,\quad t>0,
\end{equation}
with a continuous weight function $\mu$; a further generalization deals with the integration with respect to a Borel measure \cite{CGSG,K08,K09}.

Evolution equations with the fractional derivative (\ref{3.11}) are widely used in physics \cite{MK1,MK2,Ma} for modeling slow relaxation and diffusion processes; in the latter, a power-like decay of the mean square displacement of a diffusive particle appears instead of the classical exponential decay. Equations with the distributed order operators (\ref{3.12})-(\ref{3.13}) describe ultraslow processes with logarithmic decay.

Considering a general operator (\ref{3.12}), it is natural to ask the following question. Under what conditions upon a nonnegative function $k\in L_1^{\text{loc}}(\mathbb R_+)$ does the operator $\DK$ possess a right inverse (a kind of a fractional integral) and produce, as a kind of a fractional derivative, equations of evolution type? The latter means, in particular, that

(A) The Cauchy problem
\begin{equation}
\label{3.14}
(\DK u)(t)=-\lambda u(t),\quad t>0;\quad u(0)=1,
\end{equation}
where $\lambda >0$, has a unique solution $u_\lambda$, infinitely differentiable for $t>0$ and completely monotone, that is $(-1)^nu_\lambda^{(n)}(t)\ge 0$ for all $t>0$, $n=0,1,2,\ldots$.

(B) The Cauchy problem
\begin{equation}
\label{3.15}
(\DK w)(t,x)=\Delta w(t,x),\quad t>0,\ x\in \Rn;\quad w(0,x)=w_0(x),
\end{equation}
where $w_0$ is a bounded globally H\"older continuous function, that is $|w_0(\xi )-w_0(\eta )|\le C|\xi -\eta |^\gamma$, $0<\gamma \le 1$, for any $\xi ,\eta \in \Rn$, has a unique bounded solution (the notion of a solution should be defined appropriately). Moreover, the equation in (\ref{3.15}) possesses a fundamental solution of the Cauchy problem, a kernel with the property of a probability density.

Note that the well-posedness of the Cauchy problem for equations with the operator $\DK$ has been established under much weaker assumptions than those needed for (A) and (B); see \cite{Gr}.

In the above special cases (A) and (B) are satisfied; see \cite{EIK,K08}. When $\DK$ is the Caputo-Djrbashian fractional derivative $\mathbb D^{(\alpha )}$, $0<\alpha <1$, then $u_\lambda (t)=E_\alpha (-\lambda t^\alpha )$ where $E_\alpha$ is the Mittag-Leffler function:
$$
E_\alpha (z)=\sum\limits_{n=0}^\infty \frac{z^n}{\Gamma (1+\alpha n)}.
$$

It is important to note some asymptotic properties of $E_\alpha$ for real arguments \cite{GKMR}. As $z\to +\infty$, $E_\alpha (z)\sim \frac{1}\alpha e^{z^{1/\alpha}}$, whch resembles the classical case $\alpha =1$ ($E_1(z)=e^z$). Meanwhile, as $z\to -\infty$,
$$
E_\alpha (z)\sim -\frac{z^{-1}}{\Gamma (1-\alpha )},
$$
so that $u_\lambda (t)\sim Ct^{-\alpha}$, $t\to \infty$. Here and below $C$ denotes various positive constants.
This slow decay property is an origin of a large variety of applications of fractional differential equations.

In the distributed order case, where $k$ is given by (\ref{3.13}) with $\mu (0)\ne 0$, we have a logarithmic decay
$$
u_\lambda (t)\sim C(\log t)^{-1},\quad t\to \infty.
$$
A more complicated choice of $\mu$ (or a more general measure instead of $\mu \,d\alpha$) leads to a diversity of possible decay patterns.

An answer to the above questions regarding conditions upon $k$ guaranteeing (A) and (B) was given in \cite{K11}. The sufficient conditions are as follows. The Laplace transform
\begin{equation*}
\K (p)=\int\limits_0^\infty e^{-pt}k(t)\,dt
\end{equation*}
should be a Stieltjes function (or, equivalently, the function $\LL (p)=p\K (p)$ should be a complete Bernstein function),
$$
\K (p)\to \infty ,\text{ as $p\to 0$};\quad \K (p)\to 0,\text{ as $p\to \infty$};
$$
$$
\LL (p)\to 0,\text{ as $p\to 0$};\quad \LL (p)\to \infty,\text{ as $p\to \infty$}.
$$

Under these conditions, $\LL (p)$ and its analytic continuation admit an integral representation \cite{SSV}
\begin{equation}
\label{3.16}
\LL (p)=\int\limits_0^\infty \frac{p}{p+t}\sigma (dt)
\end{equation}
where $\sigma$ is a Borel measure on $[0,\infty )$, such that $\int\limits_0^\infty (1+t)^{-1}\sigma(dt) <\infty$.

{\bf 2.3.} Solutions of the evolution equations
\begin{equation}
\label{3}
\frac{\partial u_1(t,x)}{\partial t}=(A_xu_1)(t,x),
\end{equation}
\begin{equation}
\label{4}
(\DK u_{(k)})(t,x)=(A_xu_{(k)})(t,x),
\end{equation}
with the same operator $A_x$ acting in the spatial variables and the same initial conditions
$$
u_1(0,x)=\xi (x),\quad u_{(k)}(0,x)=\xi (x),
$$
typically satisfy the subordination identity, that is there exists a nonnegative function $G(s,t)$, $s,t>0$, such that $\int\limits_0^\infty G(s,t)\,ds=1$ and
\begin{equation}
\label{5}
u_{(k)}(t,x)=\int\limits_0^\infty G(s,t)u_1(s,x)\,ds.
\end{equation}

The appropriate notions of solutions of (\ref{3}) and (\ref{4}) depend on the specific setting and were explained in \cite{K11} for the case where $A$ is the Laplace operator on $\Rn$, in \cite{B1,B2,B3} (for special classes of functions $k$) in the setting with abstract semigroup generators, in \cite{Pr} for abstract Volterra equations. There is also a probabilistic interpretation of subordination identities (see, for example, \cite{Ko,Sa}). In the models of statistical dynamics  considered below,
we will deal with a subordination of measure flows  that will give a weak solution to  corresponding fractional equation.

In the above relation (\ref{5}), the subordination kernel does not depend on $A$ and can be found as follows \cite{K11}. Consider the function
$$
g(s,p)=\K (p)e^{-s\LL (p)},\quad s>0,p>0.
$$
The function $p\mapsto e^{-s\LL (p)}$ is completely  monotone (see conditions for the complete monotonicity in Chapter 13 of \cite{Fe}). By Bernstein's theorem, for each $s\ge 0$, there exists such a probability measure $\mu_s(d\tau )$ that
\begin{equation*}
e^{-s\LL (p)}=\int\limits_0^\infty e^{-p\tau }\mu_s(d\tau ).
\end{equation*}
The family of measures $\{ \mu_s\}$ is weakly continuous in $s$. Then we set
\begin{equation}
\label{6}
G(s,t)=\int\limits_0^tk(t-\tau )\mu_s(d\tau ).
\end{equation}
We can find the Laplace transform of $G$ in the variable $t$:
\begin{equation}
\label{7}
g(s,p)=\int\limits_0^\infty e^{-pt}G(s,t)\,dt.
\end{equation}

\section{Statistical dynamics and fractional kinetics}

We will consider  Markov dynamics of interacting particle systems in $\mathbb{R}^d$.
The phase space of such systems is the configuration space over the space ${{\mathbb{R}}^{d}}$  which consists
of all locally finite subsets (configurations) of ${{\mathbb{R}}^{d}}$,
namely,
\begin{equation}
\Gamma=\Gamma({{\mathbb{R}}^{d}}):=\big\{\gamma\subset{{\mathbb{R}}^{d}}\big||\gamma\cap\Lambda|<\infty,\text{ for all }\Lambda\in{\mathcal{B}}_{\mathrm{b}}({{\mathbb{R}}^{d}})\big\},\label{eq:conf_space}
\end{equation}
where ${\mathcal{B}}_{\mathrm{b}}({{\mathbb{R}}^{d}})$ denotes the family of bounded Borel subsets from ${{\mathbb{R}}^{d}}$.
The space $\Gamma$ is equipped with the vague topology, i.e., the
minimal topology for which all mappings $\Gamma\ni\gamma\mapsto\sum_{x\in\gamma}f(x)\in{\mathbb{R}}$
are continuous for any continuous function $f$ on ${{\mathbb{R}}^{d}}$
with compact support. Note that the summation in $\sum_{x\in\gamma}f(x)$
is taken over only finitely many points of $\gamma$ belonging
to the support of $f$. It was shown in \cite{Kondratiev2006} that
with the vague topology $\Gamma$ may be metrizable and it becomes
a Polish space (i.e., a complete separable metric space). Corresponding
to this topology, the Borel $\sigma$-algebra ${\mathcal{B}}(\Gamma)$
is the smallest $\sigma$-algebra for which all mappings
\[
\Gamma\ni\gamma\mapsto|\gamma_{\Lambda}|\in{\mathbb{N}}_{0}:={\mathbb{N}}\cup\{0\}
\]
 are measurable for any $\Lambda\in{\mathcal{B}}_{b}({{\mathbb{R}}^{d}})$.
Here $\gamma_{\Lambda}:=\gamma\cap\Lambda$, and $|\cdot|$ the cardinality of a finite set.
Together with $\Gamma$, it is useful to introduce a space $\Gamma_0$ which consists of all finite configurations
in ${{\mathbb{R}}^{d}}$ \cite{KK99}.

A description of each particular model  includes a heuristic Markov generator  $L$
defined on functions over the configuration space  $\Gamma$ of the system.
We assume
that the initial distribution (the state of particles) in our system is
a probability measure $\mu_{0}\in\mathcal{M}^{1}(\Gamma)$ with corresponding
sequence of correlation functions $\varkappa_{0}=(\vk_{0}^{(n)})_{n=0}^{\infty}$, see e.g.
\cite{KK99} . The distribution
of particles at time $t>0$ is the measure $\mu_{t}\in\mathcal{M}^{1}(\Gamma)$,
and $\vk_{t}=(\vk_{t}^{(n)})_{n=0}^{\infty}$ its correlation functions.
If the evolution of states $(\mu_{t})_{t\geq0}$ is determined
by a heuristic Markov generator $L$, then $\mu_{t}$ is the solution
of the forward Kolmogorov equation (or Fokker-Plank equation (FPE)),
\begin{equation}
\begin{cases}
\frac{\partial\mu_{t}}{\partial t} & =L^{*}\mu_{t}\\
\mu_{t}|_{t=0} & =\mu_{0},
\end{cases}\label{eq:FPe}
\end{equation}
where $L^{*}$ is the adjoint operator.
In terms of the time-dependent correlation functions $(\vk_{t})_{t\geq0}$
corresponding to $(\mu_{t})_{t\geq0}$, the FPE may be rewritten
as an infinite system of evolution equations
\begin{equation}
\begin{cases}
\frac{\partial \vk_{t}^{(n)}}{\partial t} & =(L^{\triangle}\vk_{t})^{(n)}\\
\vk_{t}^{(n)}|_{t=0} & =\vk_{0}^{(n)},\qquad n\geq0,
\end{cases}\label{eq:hierarchy}
\end{equation}
where $L^{\triangle}$ is the image of $L^{*}$ in a  space
of vector-functions $\vk_{t}=(\vk_{t}^{(n)})_{n=0}^{\infty}$. In applications
to concrete models, the expression for the operator $L^{\triangle}$
is obtained from the operator $L$
via
combinatoric
calculations (cf.~\cite{KK99}).

The evolution equation (\ref{eq:hierarchy}) is nothing but a hierarchical
system of equations corresponding  to the Markov generator $L$. This system is the
analogue of the BBGKY-hierarchy of the Hamiltonian dynamics \cite{Bo62}.

Our interest now turns to Vlasov-type scaling of stochastic dynamics for
the IPS in a continuum. This scaling leads to
so-called kinetic description of the considered model. In  the language of theoretical physics
we are dealing with a mean-field type scaling
which is adopted to preserve the spatial structure. In addition, this scaling will lead
to the limiting hierarchy, which possesses a chaos
preservation property. In other words, if the initial distribution
is Poisson (non-homogeneous) then the time evolution of states will
maintain this property.  We refer to \cite{FKK10} for a
general approach, concrete examples, and additional references.

There exists a standard procedure for deriving Vlasov scaling from the generator in (\ref{eq:hierarchy}).
The specific type of scaling is dictated by the model in question.
The process leading from $L^{\triangle}$ to  the rescaled Vlasov operator $L_{V}^{\triangle}$ produces a non-Markovian
generator $L_V$ since it lacks the positivity-preserving property.
Therefore instead of (\ref{eq:FPe}) we consider the following kinetic
FPE,
\begin{equation}
\begin{cases}
\frac{\partial\mu_{t}}{\partial t} & =L_{V}^{*}\mu_{t}\\
\mu_{t}|_{t =0} & =\mu_{0},
\end{cases}\label{eq:FPe1}
\end{equation}
and observe that if the initial distribution satisfies
$\mu_{0}=\pi_{\rho_{0}}$,
then the solution is of the same type, i.e., $\mu_{t}=\pi_{\rho_{t}}$.

In terms of correlation functions, the kinetic FPE (\ref{eq:FPe1})
gives rise to the following Vlasov-type hierarchical chain (Vlasov hierarchy)
\begin{equation}
\begin{cases}
\frac{\partial \vk_{t}^{(n)}}{\partial t} & =(L_{V}^{\triangle}\vk_{t})^{(n)}\\
\vk_{t}^{(n)}|_{t=0} & =\vk_{0}^{(n)},\qquad n \geq 0.
\end{cases}\label{eq:vlasov_hierarchy}
\end{equation}

Let us consider so-called Lebesgue-Poisson exponents
$$\vk_{0}(\eta)=e_{\lambda}(\rho_{0},\eta)=\prod_{x\in\eta}\rho_{0}(x)$$ as the initial condition.
Such correlation functions correspond to Poisson measures $\pi_{\rho_{0}}$ on $\Gamma$
 with the density
$\rho_{0}$.
The scaling $L_{V}^{\triangle}$ should be such  that the dynamics $\vk_{0}\mapsto \vk_{t}$
preserves this structure, or more precisely, $\vk_{t}$ should be of the same type
\begin{equation}
\vk_{t}(\eta)=e_{\lambda}(\rho_{t},\eta)=\prod_{x\in\eta}\rho_{t}(x),\quad\eta\in\Gamma_{0}.\label{eq:chaotic_preservation}
\end{equation}

 Relation (\ref{eq:chaotic_preservation}) is known as the \emph{chaos
propagation property} of the Vlasov hierarchy. It turns out that
equation (\ref{eq:chaotic_preservation}) implies, in general, a non-linear
differential equation
\begin{equation}
\frac{\partial\rho_{t}(x)}{\partial t}=\vartheta(\rho_{t})(x),\quad x\in\mathbb{R}^{d},\label{eq:nl-diff-eq-density}
\end{equation}
for $\rho_{t}$,
which is called the \emph{Vlasov-type kinetic equation}.

In general, if one does not start with a Poisson measure, the solution will leave the space $\mathcal{M}^{1}(\Gamma)$. To have a
bigger class of initial measures, we may consider
the cone inside  $\mathcal{M}^{1}(\Gamma)$
generated by convex combinations of Poisson measures, denoted by $\mathbb{P}(\Gamma)$.

Below we discuss the concept of a fractional Fokker-Plank equation  and the
related fractional statistical dynamics, which is still an evolution
in the space of probability measures on the configuration space.
The mesoscopic scaling of this evolutions leads to a fractional kinetic FPE. A subordination principle provides for the representation of
the solution to this equation as a flow of measures that is a transformation
of a Poisson flow for the initial kinetic FPE.

We will introduce the fractional statistical dynamics for a given
Markov generator $L$ by changing the time derivative in the FPE to $\DK$.
The resulting fractional Fokker-Planck dynamics (if it exists) will
act in the space of states on $\Gamma$, i.e., it
will preserve probability measures on $\Gamma$. The fractional
Fokker-Planck equation (FFPE)
\[
\begin{cases}
\DK \mu_{t}^{k} & =L^{*}\mu_{t}^{(k)}\\
\mu_{t}^{(k)}|_{t=0} & =\mu_{0}^{(k)}.
\end{cases}\tag{FFPE}
\]
describes a dynamical system with memory in the space of measures
on $\Gamma$. The corresponding evolution no longer has the semigroup
property. However, if the solution $\mu_{t}$ of equation (\ref{eq:FPe1})
exists, then the subordination principle described above shall give for the solution of (FFPE)
\begin{equation}
\mu_{t}^{(k)}=\int_{0}^{\infty} G(s,t)\mu_{s}\,ds.\label{eq:subordination1}
\end{equation}
An application of the subordination principle may be justified in many particular models where
the evolution of correlation functions may be constructed by means a $C_0$-semigroup in a proper
Banach space. In general, the subordination formula  may be considered as a rule for the transformation
of Markov dynamics to fractional ones.

It is easy to see that  $\mu_{t}^{(k)}$ is a measure.
The FFPE equation may be written in terms of time-dependent correlation
functions as an infinite system of evolution equations, the so-called
\emph{hierarchical chain}:
\[
\begin{cases}
\DK \vk_{(k),t}^{(n)} & =(L^{\triangle}\vk_{(k),t})^{(n)}\\
\vk_{(k),t}^{(n)}|_{t=0} & =\vk_{(k),0}^{(n)},\qquad n\geq0.
\end{cases}
\]
The evolution of the correlation functions should also be given by the
subordination principle. More precisely, if the solution $\vk_{t}$
of equation (\ref{eq:vlasov_hierarchy}) exists and satisfy certain exponential growth bound (as in examples considered below), then we have
\[
\vk_{(k),t}=\int_{0}^{\infty}G(s,t) \vk_{s}\,ds.
\]


As in the case of Markov statistical dynamics addressed above, we may consider Vlasov-type scaling in the framework of the FFPE.
We know that the kinetic statistical dynamics for a Poisson initial
state $\pi_{\rho_0}$ is given by a flow of Poisson measures
\[
\mathbb{R}_{+}\ni t\mapsto\mu_{t}=\pi_{\rho_{t}}\in\mathcal{M}^{1}(\Gamma),
\]
where $\rho_t$ is the solution to the corresponding Vlasov kinetic
equation. Then the fractional kinetic dynamics of states may be
obtained as the subordination of this flow.  Specifically,
 we consider the subordinated flow
\[
\mu_{t}^{(k)}:=\int_{0}^{\infty}G(s,t)\mu_{s}\,ds.
\]
The family of measures $ \mu_{t}^{(k)}$ is
no longer
a Poisson flow. We would like to analyze the properties of these subordinated flows to distinguish the
effects of fractional evolution.
It is reasonable to study the properties of subordinated flows from a more general point
of view when the evolution of densities $\rho_t(x)$ is not necessarily
related to a particular Vlasov-type kinetic  equation.

\section{Subordination and intermittency}

 As we already discussed,  for the fractional kinetic hierarchies  the correlation functions have the following representation
$$
\varkappa^{(n)}_t (x_1,\ldots ,x_n)=\int\limits_0^\infty G(s,t)\prod\limits_{j=1}^n \varkappa_s^{(1)}(x_j)\,ds.
$$
 Let us consider  a model situation $\varkappa_s^{(1)}(x)\equiv e^{\beta s}$, $\beta >0$, so that
\begin{equation}
\label{8}
\varkappa^{(n)}_t=\int\limits_0^\infty G(s,t)e^{n\beta s}\,ds.
\end{equation}
This situation is realized, in particular, in the kinetic limit of the spatial contact model in the supercritical regime, see \cite{FKK10, Kondratiev2008}.
The existence of the integral in (\ref{8}) will be proved later.  We will study an intermittency property  of the solution to the kinetic hierarchy in the considered case. For a general discussion concerning the notion of intermittency see
\cite{Carmona1994, Carmona1995}.  Note that the intermittency property for random fields are formulated in terms of their moments. But in the case of random point processes we are dealing with there is an alternative possibility to reformulate this property in terms of correlation functions, see
\cite{DaSilvaKochKon}.
For our case, the intermittency property means that for each $n>1$, and the natural numbers $m_1,\ldots ,m_k$, such that $m_1+\cdots +m_k=n$,
\begin{equation}
\label{9}
\frac{\varkappa^{(n)}_t}{\prod\limits_{j=1}^k \varkappa^{(m_j)}_t}\longrightarrow \infty,\quad \text{as $t\to \infty$}.
\end{equation}

\medskip
\begin{teo}
The intermittency property (\ref{9}) is fulfilled, if
\begin{equation}
\label{10}
\int\limits_1^\infty \frac{ds}{s\LL (s)}<\infty .
\end{equation}
\end{teo}

\medskip
{\it Proof}. Let us consider the function
\begin{equation}
\label{11}
A(t,z)=\int\limits_0^\infty e^{zs}G(s,t)\,ds,\quad t>0,z>0.
\end{equation}
By the Fubini-Tonelli theorem, the existence of the integral in (\ref{11}) for almost all $t>0$ follows from the absolute convergence of the repeated integral
$$
\int\limits_0^\infty e^{zs}\,ds\int\limits_0^\infty e^{-pt}G(s,t)\,dt=\int\limits_0^\infty e^{zs}g(s,p)\,ds=\frac{\K (p)}{\LL (p)-z}
$$
where $p>0$ is such that $\LL (p)>z$.

Therefore the function $A(t,z)$ exists almost everywhere and is locally integrable in $t$ for each fixed $z$. Its Laplace transform
$$
\A (p,z)=\int\limits_0^\infty e^{-pt}A(t,z)\,dt
$$
is defined for $\LL (p)>z$. For such values of $p$,
\begin{equation*}
\A (p,z)=\frac{\K (p)}{\LL (p)-z}.
\end{equation*}

Since $\LL (p)$ is a Bernstein function, that is its derivative is completely monotone, by Bernstein's theorem, $\LL'(p)\ne 0$ for all $p>0$ ($\LL (p)$ is not a constant function by our assumptions). Therefore $\LL$ is strictly monotone. For each $z>0$, there exists a unique $p_0=p_0(z)$, such that $\LL (p_0)=z$. The condition $\LL (p)>z$ is equivalent to the inequality $p>p_0(z)$. Note that, by virtue of (\ref{3.16}), $\LL (p)\ne z$ for any nonreal $p$, since $\LL (p)$ preserves the open upper and lower half-planes.

It follows from (\ref{3.16}) that $\A (p,z)$ is holomorphic in $p$ on any sector $p_0+\Sigma_{\rho +\frac{\pi}2}$, $0<\rho <\frac{\pi}2$. Here $\Sigma_\delta =\{ re^{i\theta}:\ r>0,-\delta <\theta <\delta \}$, $\delta >0$. In addition,
$$
\sup\limits_{p\in p_0+\Sigma_{\rho +\frac{\pi}2}}|(p-p_0)\A (p,z)|<\infty .
$$
By Theorem 2.6.1 from \cite{ABHN}, the function $A(t,z)$ is actually holomorphic in $t$ on any sector $\Sigma_\upsilon$, $0<\upsilon <\rho$, and
$$
\sup\limits_{t\in \Sigma_\upsilon}\left| e^{-p_0t}A(t,z)\right|<\infty .
$$

Let us rewrite $\A$ as follows:
$$
\A (p,z)=\frac1p\left( 1+\frac{z}{\LL(p)-z}\right) .
$$
This implies the relation $A(t,z)=1+B(t,z)$ where $B$ has the Laplace transform
$$
\B (p,z)=\frac{z}p\cdot \frac1{\LL (p)-z}.
$$

It is known that the complete Bernstein function $\LL$ satisfies, outside the negative real semi-axis, the inequality
\begin{equation}
\label{12}
\sqrt{\frac{1+\cos \varphi}2}\LL (|p|)\le |\LL (p)|\le \sqrt{\frac2{1+\cos \varphi}}\LL (|p|),\quad \varphi =\arg p
\end{equation}
(see Proposition 2.4 in \cite{BCT}). In particular, on any vertical line $p=\gamma +i\lambda$, $\gamma >p_0$, we have $|\LL (p)|\ge \frac1{\sqrt{2}}\LL (|p|)$. Together with the assumption (\ref{10}), this implies the absolute integrability on such a line of the function $\B (p,z)$, as well as the fact that $\B (p,z)\to 0$, as $p\to \infty$ in the half-plane $\R p>p_0$. Having these properties (see Theorem 28.2 in \cite{Do}), we can write the inversion formula
\begin{equation}
\label{13}
A(t,z)=1+\frac{z}{2\pi i}\int\limits_{\gamma -i\infty}^{\gamma +i\infty} e^{pt}\frac{dp}{p(\LL (p)-z)},\quad \gamma >p_0.
\end{equation}

We use (\ref{13}) to study the asymptotics of $A(t,z)$ for a fixed $z$, as $t\to \infty$. Denote
$$
I_0(t,z)=1+\frac{z}{2\pi i}\int\limits_{r-i\infty}^{r+i\infty} e^{pt}\frac{dp}{p(\LL (p)-z)}
$$
where $0<r<p_0$. We have
\begin{equation}
\label{14}
\left| I_0(t,z)\right| \le 1+Ce^{rt}\left| \int\limits_{-\infty}^{\infty}e^{i\lambda t}\frac{d\lambda}{(r+i\lambda )(\LL (r+i\lambda )-z)}\right| =o(e^{rt}),\quad t\to \infty,
\end{equation}
due to (\ref{10}), (\ref{12}) and the Riemann-Lebesgue theorem.

On the other hand,
$$
A(t,z)-I_0(t,z)=\frac{z}{2\pi i}\left( \int\limits_{\Gamma_+}+\int\limits_{\Gamma_0}+\int\limits_{\Gamma_-}\right) e^{pt}\frac{dp}{p(\LL (p)-z)}
$$
where the contour $\Gamma_+$ consists of the vertical rays $\{ \R p=r,\I p\ge R\}$, $\{ \R p=\gamma,\I p\ge R\}$, and the horizontal segment $\{ r\le \R p\le \gamma, \I p=R\}$ ($R>0$), $\Gamma_-$ is a mirror reflection of $\Gamma_+$ with respect to the real axis, $\Gamma_0$ is the finite rectangular contour consisting of the vertical segments $\{ \R p=r,|\I p|\le R\}$, $\{ \R p=\gamma ,|\I p|\le R\}$, and the horizontal segments $\{ r\le \R p\le \gamma,\I p=\pm R\}$.

We have
$$
\int\limits_{\Gamma_+}e^{pt}\frac{dp}{p(\LL (p)-z)}=0.
$$

That follows from the Cauchy theorem, absolute integrability of the integrand on the vertical rays, and the estimate of the integral over the horizontal segment $\Pi_h=\{ r\le \R p\le \gamma,\I p=h\}$ ($h>R)$:
$$
\left| \int\limits_{\Pi_h}e^{pt}\frac{dp}{p(\LL (p)-z)}\right| \le Ch^{-1}\to 0,
$$
as $h\to \infty$. Similarly,
$$
\int\limits_{\Gamma_-}e^{pt}\frac{dp}{p(\LL (p)-z)}=0.
$$

Since $\LL'(p_0)\ne 0$, there exists a complex neighborhood $U$ of $z=\LL (p_0)$, where the function $\LL$ possesses a single-valued holomorphic inverse function $p=\psi (w)$, so that $\LL (\psi (w))=w$ and $p_0=\psi (z)$. Up to now, the numbers $r,\gamma,R$ were arbitrary. Choose $R$ and $\gamma -r$ so small that the curvilinear rectangle $\LL (\Gamma_0)$ lies within $U$. Making the change of variables $p=\psi (w)$ and using the Cauchy formula we find that
\begin{multline*}
\frac{z}{2\pi i}\int\limits_{\Gamma_0}e^{pt}\frac{dp}{p(\LL (p)-z)}=\frac{z}{2\pi i}\int\limits_{\LL (\Gamma_0)}e^{\psi (w)t}\frac1{\LL'(\psi (w))\psi (w)}\cdot \frac{dw}{w-z}
=\frac{z}{\LL'(\psi (z))\psi (z)}e^{\psi (z)t}\\
=\frac{z}{\LL'(p_0(z))p_0(z)}e^{p_0(z)t}.
\end{multline*}

Together with (\ref{14}), this yields the asymptotic relation
\begin{equation}
\label{15}
A(t,z)=\frac{z}{\LL'(p_0(z))p_0(z)}e^{p_0(z)t}+o(e^{p_0(z)t}),\quad t\to \infty.
\end{equation}

\medskip
In the next lemma we use the duality of sub- and superadditivity \cite{Os}.

\medskip
\begin{lem}
The function $p_0(z)$, $z>0$, is strictly superadditive; in particular,
$$
p_0(n\beta )>\sum\limits_{j=1}^k p_0(m_j\beta )
$$
for $n=\sum\limits_{j=1}^k m_j$, $\beta >0$.
\end{lem}

\medskip
{\it Proof}. It is sufficient to prove that
\begin{equation}
\label{4.*}
p_0(x+y)>p_0(x)+p_0(y)\quad \text{for any $x,y>0$}.
\end{equation}

First of all, $\LL$ is strictly subadditive, that is
\begin{equation}
\label{4.**}
\LL (a+b)<\LL (a)+\LL (b),\quad a,b>0.
\end{equation}
This follows from the integral representation (\ref{3.16}) and the elementary identity
$$
\frac{a}{a+t}+\frac{b}{b+t}-\frac{a+b}{a+b+t}=\frac{a^2b+2abt+ab^2}{(a+t)(b+t)(a+b+t)}
$$
We get from (\ref{4.**}) and the strict monotonicity of $p_0$ that
$$
a+b<p_0(\LL (a)+\LL (b)).
$$

By our assumptions, $\LL$ maps bijectively the semi-axis $[0,\infty )$ onto itself. Choosing $a,b$ in such a way that $\LL(a)=x$, $\LL (b)=y$, so that $a=p_0(x)$, $b=p_0(y)$, we obtain (\ref{4.*}). $\qquad \blacksquare$

\medskip
The asymptotic relation (\ref{10}) follows from  (\ref{15}) and Lemma 4.1. $\qquad \blacksquare$

\bigskip
{\it Examples}. 1). For the Caputo-Djrbashian fractional derivative $\D$, $0<\alpha <1$, we have $\LL (p)=p^\alpha$, so that (\ref{10}) is satisfied.

Note that \cite{B1} in this case $A(t,z)=E_\alpha (zt^\alpha )$ where $E_\alpha$ is the Mittag-Leffler function, and the asymptotic relation (\ref{15}) gives actually the principal term of the asymptotics of $E_\alpha$. However our proof is different from the well-known proof of the latter (see \cite{Dj,GKMR}).

\medskip
2). Consider a distributed order derivative with a continuous weight function $\mu$, that is
$$
\mathbb D^{(\mu )}\varphi (t)=\int\limits_0^1 (\mathbb D^{(\alpha )}\varphi )(t)\mu (\alpha )\,d\alpha .
$$
In this case (see \cite{K08}),
$$
k(s)=\int\limits_0^1\frac{s^{-\alpha}}{\Gamma (1-\alpha)}\mu (\alpha )\,d\alpha,\quad \LL (p)=\int\limits_0^1 p^\alpha \mu (\alpha )\,d\alpha .
$$
It is proved in \cite{K08} that, if $\mu \in C^2[0,1]$, then
$$
\LL (p)=\frac{\mu (1)p}{\log p}+O(p|\log p|^{-2}),\quad p\to \infty .
$$
Therefore (\ref{10}) is satisfied, if $\mu (1)\ne 0$. See \cite{K08} for an investigation of the case where $\mu (1)=0$.

\bigskip
 In the model with decaying correlation functions, it is assumed that
\begin{equation}
\label{16}
\varkappa_t^{(1)}=e^{-\beta t},\quad \beta >0.
\end{equation}
This situation is realized in the contact model in subcritical regime \cite{Kondratiev2008}.
\medskip
\begin{teo}
The intermittency property (\ref{9}) is fulfilled in the case (\ref{16}), if
\begin{equation}
\label{17}
\K (p)\sim p^{-\gamma}Q\left(\frac1p\right),\quad p\to 0,
\end{equation}
where $0\le \gamma \le 1$, $Q$ is a slowly varying function \cite{Fe,Se}.
\end{teo}

\medskip
{\it Proof}. Consider the function
$$
A(t,-z)=\int\limits_0^\infty e^{-zs}G(s,t)\,ds,\quad t>0,z>0.
$$
The existence and boundedness ($\le 1$) of this function is obvious. As in the above case, it is in fact analytic in $t$. Its Laplace transform
$$
\A(p,-z)=\frac{\K (p)}{\LL (p)+z}
$$
is a Stieltjes function in the variable $p$ because $p\A(p,-z)=\dfrac{\LL (p)}{\LL (p)+z}$ is a complete Bernstein function as a composition of the functions $\LL$ and $p\mapsto \dfrac{p}{p+z}$ belonging to this class.

Under our assumptions,
$$
\A(p,-z)=\int\limits_0^\infty \frac{\sigma (dr)}{p+r},
$$
where $\sigma$ is a Borel measure, $\int\limits_0^\infty \dfrac{\sigma (dr)}{1+r}<\infty$. To simplify notations, we will write temporarily $h(p)$ instead of $\A(p,-z)$ (with a fixed $z$).

\medskip
\begin{lem}
For each $n\ge 1$, $p>0$,
\begin{equation}
\label{18}
\left| \frac{h^{(n+1)}(p)}{h^{(n)}(p)}\right| \le \frac{n+1}p.
\end{equation}
\end{lem}

\medskip
{\it Proof}. We have
$$
h^{(n)}(p)=(-1)^n n!\int\limits_0^\infty \frac{\sigma (dr)}{(p+r)^{n+1}},\quad n=1,2,\ldots ,
$$
so that
$$
\left| h^{(n+1)}(p)\right| =\int\limits_0^\infty \frac{(n+1)!}{(p+r)^{n+2}}\sigma (dr)=\int\limits_0^\infty \frac{n!}{(p+r)^{n+1}}\cdot \frac{n+1}{r+p}\sigma (dr)
\le \frac{n+1}p\left| h^{(n)}(p)\right|,
$$
which implies (\ref{18}). $\qquad \blacksquare$

\medskip
A similar inequality was proved for complete Bernstein functions in \cite{Ja} (Lemma 3.9.34).

\medskip
{\it Proof of Theorem 2 (continued)}. By the Post-Widder formula (see, for example, Theorem 3.8.6 in \cite{Ja}), for a fixed $z$,
$$
A(t,-z)=\lim\limits_{n\to \infty}H_n(t),\quad t>0,
$$
where
$$
H_n(t)=\frac{(-1)^n}{n!}h^{(n)}\left(\frac{n}t\right) \left(\frac{n}t\right)^{n+1}.
$$

Each function $H_n$ is non-increasing. Indeed,
\begin{multline*}
H_n'(t)=\frac{(-1)^n}{n!}\left( -\frac{n}{t^2}\right) h^{(n+1)}\left(\frac{n}t\right) \left(\frac{n}t\right)^{n+1}+\frac{(-1)^n}{n!}h^{(n)}\left(\frac{n}t\right) (n+1)\left( -\frac{n}{t^2}\right)\left(\frac{n}t\right)^n\\
=\frac{(-1)^n}{n!}\left(\frac{n}t\right)^{n+1}\left( -\frac{n}{t^2}\right)h^{(n)}\left(\frac{n}t\right)\left[ \frac{h^{(n+1)}\left(\frac{n}t\right)}{h^{(n)}\left(\frac{n}t\right)}+\frac{(n+1)t}n\right].
\end{multline*}
By (\ref{18}), the expression in square brackets is nonnegative. Since a Stieltjes function is completely monotone, we find that $H_n'(t)\le 0$ for each $n$, that is $A(t,-z)$ is non-increasing in the variable $t$.

By the Karamata-Feller Tauberian theorem \cite{Fe}, the asymptotics (\ref{17}) and the above monotonicity of $A$ imply the asymptotics
$$
A(t,-z)\sim \frac1{\Gamma (\gamma )z}t^{\gamma -1}Q(t),\quad t\to \infty .
$$
In particular, $\varkappa_t^{(n)}=A(t,-n\beta )$, so that
$$
\frac{\varkappa^{(n)}_t}{\prod\limits_{j=1}^k \varkappa^{(m_j)}_t}\sim \operatorname{const}\cdot e^{n\beta t}t^{\gamma -1}Q(t)\longrightarrow \infty ,
$$
as $t\to \infty$. $\qquad \blacksquare$

\bigskip
{\it Examples}. 1) In the case of the Caputo-Djrbashian fractional derivative $\D$, $0<\alpha <1$, we have $\K (p)=p^{\alpha -1}$, and (\ref{17}) is satisfied.

2) For the distributed order derivative with a continuous weight function $\mu$, we have
$$
\K (p)\sim p^{-1}\left(\log \frac1p \right)^{-1}\mu (0),\quad p\to 0,
$$
if $\mu (0)\ne 0$. Thus, in this case (\ref{17}) holds with $\gamma =1$.

\section*{Acknowledgment}
This paper was partially supported by the DFG through the SFB 701
``Spektrale Strukturen und Topologische Methoden in der Mathematik''
and by the European Commission under the project STREVCOMS
PIRSES-2013-612669.

\end{document}